\documentstyle[12pt,epsfig]{article}
\begin{document}
\noindent
\begin{center}
{\Large {\bf Cosmic Acceleration in Brans-Dicke Cosmology}}\\ \vspace{2cm}
 ${\bf Yousef~Bisabr}$\footnote{e-mail:~y-bisabr@srttu.edu.}\\
\vspace{.5cm} {\small{Department of Physics, Shahid Rajaee Teacher
Training University,
Lavizan, Tehran 16788, Iran}}\\
\end{center}
\vspace{1cm}
\begin{abstract}
We consider Brans-Dicke theory with a self-interacting potential in Einstein conformal frame.  We show that an accelerating expansion
is possible in a spatially flat universe for large values of the Brans-Dicke parameter consistent with local gravity experiments.

\end{abstract}
\vspace{3cm}
Recent observations
on expansion history of the universe indicate
that the universe is experiencing a phase of accelerated expansion
\cite{sup}.  There are two classes of models in the literature for explaining these observations.  In one class, cosmic
acceleration is attributed to some exotic matter components with a negative
pressure, dubbed dark energy \cite{mel}. This class also includes quintessence models \cite{q}, a minimally coupled scalar field
with a potential which can generate a sufficiently negative pressure at the present epoch.  It should be pointed out that in an observational
point of view, models with
a cosmological term or adiabatic $\Lambda$CDM (cosmological constant or non-evolving dark energy) seem to be in agreement with cosmological
observations \cite{ref}.  However, the observational
upper bound differs from theoretical
expectations by more than 100 orders
of magnitude \cite{cc}.  This large discrepancy avoids people to consider the cosmological constant as a viable candidate
of dark energy \cite{bis}.\\
 Alternatively, there is a class of models
that propose a modification of the gravitational part of the
Einstein-Hilbert action.  This includes scalar-tensor theories, scalar fields non-minimally coupled to gravity.
Here we shall consider a self-interacting Brans-Dicke (BD) theory \cite{BD} as a prototype of scalar-tensor theories.  The
original motivation of the BD theory was the search for a theory containing
Mach's principle which has found a limited expression in general relativity.  As
the simplest and best-studied generalization of general relativity, it is natural to think
about the BD scalar field as a possible candidate for producing cosmic acceleration without invoking a quintessence field or
 exotic matter systems.  In fact, there have been many
attempts to show that BD model can potentially
explain the cosmic acceleration.  It is shown that this theory can actually produce a non-decelerating
expansion for low negative values of the BD
 parameter $\omega$ \cite{ban}. This conflicts with the lower
bound imposed on this parameter by solar system experiments \cite{will}. Some authors
propose modifications of the BD model such as introducing a potential function for the scalar field \cite{ban1}, or
considering a field-dependent BD parameter \cite{ban2}, without resolving this problem.\\
All the works in this context use Jordan frame representation of BD theory.  It is however well-known
that this theory, like any other scalar-tensor theories, can be represented in the so-called Einstein frame
by using a conformal transformation \cite{far1} \cite{far}.  Although these two conformal frames are mathematically
equivalent there are some debates on their physical equivalence.  Here there is a point which should be made clear.  Despite the fact that the problem of physical status of the two conformal
frames is open there is a tendency in the literature to ignore this problem
and to work in Jordan conformal frame. The reason may be related to reluctance in accepting of the violation of weak equivalence principle
due to anomalous coupling of the scalar field to matter systems in Einstein frame.  It is however important to keep in mind that
the physical metric should be singled out already in the vacuum sector of the theory and the coupling of a given metric
to matter systems is determined by the physical significance ascribed to it \cite{soko}.  Thus a
criterion based on the coupling of matter with gravity would be effective only if the physical frame were
determined on an independent ground.  Apart from this point, anomalous gravitational coupling in Einstein conformal frame does not
necessarily mean violation of weak equivalence principle.  There is still a possibility
that the effective mass of the scalar field be scale dependent. In this chameleon mechanism \cite{ch}, the scalar field may
acquire a large effective mass in Solar System scale so that it hides
local experiments while at cosmological scales it is effectively light
and can provide an appropriate cosmological behavior. \\ Along these lines, we would like to consider Einstein frame formulation of the theory as a representation
which provides different possibilities in a cosmological setting with respect to the Jordan frame.  In particular, we will focus on the question that whether it is possible to achieve accelerating
expansion of the universe for sufficiently large values of the parameter $\omega$.  We will show that the answer is affirmative within a class of solutions which
corresponds to a specific form of the potential function of the BD scalar field.\\\\
We begin with a modified form of the BD action in Jordan frame
\begin{equation}
S_{JF}= \int d^{4}x \sqrt{-\bar{g}} (\phi \bar{R} -\frac{\omega}{\phi}\bar{g}^{\mu\nu}\bar{\nabla}_{\mu}\phi \bar{\nabla}_{\nu}\phi-V(\phi))+S_{m}(\bar{g}_{\mu\nu}, \psi)\label{a1}\end{equation}
where $\phi$ is the self-interacting BD scalar field with a potential function $V(\phi)$, $\omega$ is a constant parameter and  $S_{m}$ is the action
of matter which depends on the metric $\bar{g}_{\mu\nu}$ and some matter
fields collectively denoted by $\psi$. A conformal transformation
\begin{equation}
\bar{g}_{\mu\nu}\rightarrow g_{\mu\nu}=\Omega^2 \bar{g}_{\mu\nu}
\label{a2}\end{equation}
with $\Omega=\sqrt{G \phi}$ brings the above action into the Einstein frame \cite{far1} \cite{far}.  Then a scalar field redefinition
\begin{equation}
\varphi(\phi)=\sqrt{\frac{2\omega+3}{16\pi G}}\ln (\frac{\phi}{\phi_0})
\label{a3}\end{equation}
with $\phi_0\sim G^{-1}$, $\phi>0$ and $\omega>-\frac{3}{2}$ transforms the kinetic term of the scalar field into a canonical form.  In terms of the variables
($g_{\mu\nu}$, $\varphi$) the BD action in the Einstein frame is \cite{far1} \cite{far}
\begin{equation}
S_{EF}= \int d^{4}x \sqrt{-g} (\frac{R}{16\pi G} -\frac{1}{2}g^{\mu\nu}\nabla_{\mu}\varphi \nabla_{\nu}\varphi-U(\varphi))+S_{m}(g_{\mu\nu}, \psi)
\label{a4}\end{equation}
where
\begin{equation}
S_{m}= \int d^{4}x \sqrt{-g}~ \exp(-8\sqrt{\frac{\pi G}{2\omega+3}}\varphi)~L_{m}(g_{\mu\nu}, \psi)
\label{a5}\end{equation}
Here $\nabla_{\mu}$ is the covariant derivative of the rescaled metric $g_{\mu\nu}$ and
\begin{equation}
U(\varphi)= V(\phi(\varphi))~\exp(-8\sqrt{\frac{\pi G}{2\omega+3}}\varphi)
\label{a6}\end{equation}
is the Einstein frame potential.   \\
Variation of the action (\ref{a4}) with respect to $g_{\mu\nu}$ and $\varphi$ leads to the following field equations
\begin{equation}
G_{\mu\nu}=8\pi G (T_{\mu\nu}+T^{\varphi}_{\mu\nu})
\label{a7}\end{equation}
\begin{equation}
\Box\varphi-\frac{dU(\varphi)}{d\varphi}=-\frac{1}{2}\alpha~T
\label{a8}\end{equation}
where
\begin{equation}
T^{\varphi}_{\mu\nu}=\nabla_{\mu}\varphi \nabla_{\nu}\varphi-\frac{1}{2}g_{\mu\nu}\nabla_{\gamma}\varphi \nabla^{\gamma}\varphi-U(\varphi)g_{\mu\nu}~,
\label{a9}\end{equation}
$\alpha = \sqrt{\frac{16\pi G}{2\omega+3}}$ and $T=g^{\mu\nu}T_{\mu\nu}$ is the trace of the matter stress-tensor.  Note that the parameter
$\alpha$ is related to inverse of the BD parameter $\omega$.  In Einstein frame, the vacuum sector of the action consists of a scalar
field minimally coupled to Einstein's gravity.  The important difference between the Einstein frame representation of BD model and minimally coupled scalar
field models is that in the former the scalar field interacts with matter systems.  This anomalous gravitational coupling has no counterpart in Einstein's gravity.  It implies that the stress-tensors of matter and the scalar field are not
separately conserved.  This can be easily checked by applying Bianchi's identities to (\ref{a7}) which leads to
\begin{equation}
\nabla^{\mu}T_{\mu\nu}=-\nabla^{\mu}T^{\varphi}_{\mu\nu}=\frac{1}{2}\alpha~T~ \nabla_{\nu}\varphi
\label{1a9}\end{equation}
The parameter $\alpha$ measures the strength of the interaction.  Here there are two important points in order :
First, the parameter $\alpha$ is positive $\alpha>0$.  It implies that energy transfer is from scalar field $\varphi$ to matter systems.  This feature is consistent with the second law of thermodynamics \cite{pa}.  Second, the model
(\ref{a4}) should be constrained by local gravity experiments to avoid violation of weak equivalence principle.  It is well-known that these constraints are satisfied when $\omega>>1$ \cite{will}.  In the model
(\ref{a4}) this translates into $\alpha<<1$.  This means that the theory can pass local tests if interaction of the scalar
field $\varphi$ with matter fields is sufficiently small.  We will
return to this issue later.\\
We apply the field equations (\ref{a7}) and (\ref{a8}) to a spatially flat Friedmann-Robertson-Walker spacetime
\begin{equation}
ds^2=-dt^2+a^2(t)(dx^2+dy^2+dz^2)
\label{a10}\end{equation}
with $a(t)$ being the scale factor.  To do this, we take the matter system to be a pressureless perfect fluid (dust) with energy
density $\rho_m$.  In this case, the gravitational equations (\ref{a7}) give
\begin{equation}
3\frac{\dot{a}^2}{a^2}=k(\rho_{m}+\rho_{\varphi})
\label{a11}\end{equation}
\begin{equation}
2\frac{\ddot{a}}{a}+\frac{\dot{a}^2}{a^2}=-k~p_{\varphi}
\label{a12}\end{equation}
where $k=8\pi G$, $\rho_{\varphi}=\frac{1}{2}\dot{\varphi}^2+U(\varphi)$ and $p_{\varphi}=\frac{1}{2}\dot{\varphi}^2-U(\varphi)$. We may use the
first equation to rewrite the second one as
\begin{equation}
\frac{\ddot{a}}{a}+2\frac{\dot{a}^2}{a^2}=\frac{1}{2}k\rho_m+kU(\varphi)
\label{a12-1}\end{equation}
The equation (\ref{a8}) gives
\begin{equation}
\ddot{\varphi}+3\frac{\dot{a}}{a}\dot{\varphi}+\frac{dU(\varphi)}{d\varphi}=-\frac{1}{2}\alpha ~\rho_{m}
\label{a13}\end{equation}
On the other hand, the conservation
equations (\ref{1a9}) become
\begin{equation}
\dot{\rho}_{m}+3\frac{\dot{a}}{a}\rho_m=\frac{1}{2}\alpha~\dot{\varphi}~\rho_m
\label{a14}\end{equation}
\begin{equation}
\dot{\rho}_{\varphi}+3\frac{\dot{a}}{a}(\omega_{\varphi}+1)\rho_{\varphi}=-\frac{1}{2}\alpha~\dot{\varphi}~\rho_m
\label{a15}\end{equation}
with $\omega_{\varphi}=p_{\varphi}/\rho_{\varphi}$ being the equation of state parameter of the scalar field $\varphi$.
The equation (\ref{a14}) can be solved which gives the following solution
\begin{equation}
\rho_ma^3=\rho_{m0}e^{\frac{1}{2}\alpha\varphi}
\label{a16}\end{equation}
where $\rho_{m0}$ is the present matter energy density in the universe.  To proceed further, we introduce an ansatz
\begin{equation}
\varphi=\frac{\beta}{\sqrt{k}}\ln a
\label{a17}\end{equation}
in which $\beta$ is a positive constant parameter of order of unity.  One of the advantages of this ansatz is that it brings the solution (\ref{a16}) into
the following form
\begin{equation}
\rho_m=\rho_{m0}a^{-3+\varepsilon}
\label{a18}\end{equation}
where $\varepsilon \equiv \alpha\beta /2\sqrt{k} = \beta [2(2\omega+3)]^{-\frac{1}{2}}>0$.  This is similar to the rule presented by some authors for
characterizing decaying law of vacuum energy into dark matter \cite{al}.  It states that the scalar field $\varphi$ is constantly decaying
into the matter so that the latter will dilute
more slowly compared to its standard evolution $\rho_m \propto a^{-3}$.  Since the observational lower bound imposed by solar system experiments
on the BD parameter is $\omega >> 1$, we should have $\varepsilon<<1$ which means that evolution of matter density
has a small deviation with respect to the standard one in Einstein's gravity.\\
If we put the ansatz (\ref{a17}) into (\ref{a13}), we obtain
\begin{equation}
\frac{\ddot{a}}{a}+2\frac{\dot{a}^2}{a^2}=-\frac{\sqrt{k}}{\beta}(\frac{dU(\varphi)}{\varphi}+\frac{1}{2}\alpha\rho_m)
\label{a18-1}\end{equation}
Comparing the latter with (\ref{a12-1}) leads to a consistency relation
\begin{equation}
\frac{dU(\varphi)}{d\varphi}+\sqrt{k}\beta U(\varphi)=-\frac{\sqrt{k}}{\beta}(\frac{1}{2}\beta^2+\varepsilon)\rho_{m0}e^{\frac{\sqrt{k}}{\beta}(\varepsilon-3)\varphi}
\label{a18-2}\end{equation}
This consistency relation will be satisfied for an appropriate potential $U(\varphi)$.  To find the form of the potential, we solve this first order differential equation which gives the following solution
\begin{equation}
U(\varphi)=-\gamma \rho_{m0}e^{\frac{\sqrt{k}}{\beta}(\varepsilon-3)\varphi}+Ce^{-\sqrt{k}\beta\varphi}
\label{a18-2}\end{equation}
where $\gamma=\frac{\frac{1}{2}\beta^2+\varepsilon}{\beta^2+\varepsilon-3}$ and $C$ is an integration constant.
Thus the relation (\ref{a17}) is a solution of the field equations for an exponential potential of the form (\ref{a18-2}).  This double
exponential potential is similar to the potential which is used in some quintessence models \cite{cop} \cite{sen}.  It is shown that this kind of potential
of the quintessence field can lead to solutions
which first enter a period of scaling through the radiation
and matter domination eras and then smoothly evolve to dominate the energy density for a wide range of initial conditions of
the field\footnote{Note that Einstein frame representation of BD models are effectively equivalent
to the so-called coupled quintessence models in which the quintessence field interacts with matter sector.}\cite{cop}.  Moreover, single exponential potentials
are popular in modified $f(R)$ gravity models \cite{farso}.  These models are conformally equivalent to BD models with potentials which their
forms are closely related to the functional form of the $f(R)$ functions \cite{y}. In that context, single exponential potentials correspond
to power law $f(R)$ gravity models \cite{yy}.\\
The integration constant $C$
can be determined by noting the fact that when we set $\phi=\phi_0 \sim G^{-1}$ in the action (\ref{a1}), then $V(\phi)$ characterizes the vacuum
energy density corresponding to a cosmological constant, namely $V(\phi)=\Lambda/G$.  In this case, $\varphi=0$ and then
\begin{equation}
U(\varphi=0)=V(\phi(\varphi))\equiv\rho_{\varphi0}
\label{a18-3}\end{equation}
with $\rho_{\varphi0}$ being the vacuum energy density in the Einstein frame.  Applying the latter condition to the
relation (\ref{a18-2}) gives $C=\rho_{\varphi0}+\gamma \rho_{m0}$.  The relation (\ref{a18-2}) takes then the form
\begin{equation}
U(\varphi)=\gamma\rho_{m0}(e^{-\sqrt{k}\beta\varphi}-e^{\frac{\sqrt{k}}{\beta}(\varepsilon-3)\varphi})+\rho_{\varphi0}e^{-\sqrt{k}\beta\varphi}
\label{a18-4}\end{equation}
For this potential function, the Friedmann equation (\ref{a11}) becomes
\begin{equation}
\frac{H^2}{H_0^2}=\frac{3}{3-\frac{1}{2}\beta^2}[(1-\gamma)\Omega_{m0}a^{-3+\varepsilon}+ (\Omega_{\varphi 0}+\gamma\Omega_{m0})a^{-\beta^2}]
\label{a19}\end{equation}
where $\Omega_{m0}=\rho_{m0}/\rho_c$, $\Omega_{\varphi 0}=\rho_{\varphi 0}/\rho_c$ and $\rho_c=3H_0^2/k$ is the critical density.  From
the equations (\ref{a12}) and (\ref{a19}), it is straightforward to show that the deceleration parameter $q=-1-\frac{\dot{H}}{H^2}$ takes
the following form
\begin{equation}
q(z)=\frac{1}{2}\{(1+\frac{1}{2}\beta^2)-(3-\frac{1}{2}\beta^2)[1-\frac{1}{\gamma-(\frac{\Omega_{\varphi0}}{\Omega_{m0}}
+\gamma)(z+1)^{-3+\varepsilon+\beta^2}}]^{-1}\}
\label{a20}\end{equation}
where we have used $a(z)=(z+1)^{-1}$.   This relation gives deceleration parameter in terms of the redshift and constant free parameters $\omega$ and $\beta$.  We plot
evolution of $q(z)$ in fig.1.
\begin{figure}[ht]
\begin{center}
\includegraphics[width=0.6\linewidth]{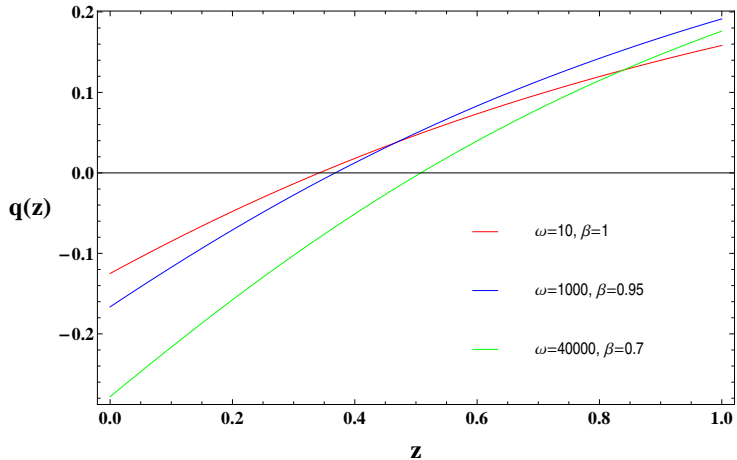}
\caption{The plot of deceleration parameter $q(z)$, given by (\ref{a20}), for some values of the parameters $\omega$ and $\beta$.  We have set
$\frac{\Omega_{m0}}{\Omega_{\varphi 0}}=\frac{3}{7}$.  The three lines correspond to $\varepsilon=0.147, 0.015, 0.002$ from top to bottom.}
\end{center}
\end{figure}
The figure shows that accelerating expansion is possible for positive small and large values of the BD parameter.
However, it should be noted
that a small $\omega$ is excluded  by two independent observations.  First, the local gravity tests
which require that $\omega>40000$ \cite{will}.  Second, from the
relation (\ref{a18}) we must have $\varepsilon<<1$ which corresponds to $\omega>>1$.  Otherwise the universe will
expand accelerated in the matter dominated era, which is
against the observation of SNe Ia that our universe
expanded decelerated before the redshift $z\sim 0.5$ \cite{r}.  To clarify this point, we consider the Friedmann equation (\ref{a11}) or
(\ref{a19}) in matter domination regime in which $\rho_{m}>>\rho_{\varphi}$.  In this case, one can simply check that
$a\propto t^{2/3-\varepsilon}$ and $H=\frac{2}{3-\varepsilon}t^{-1}$. Only for $\varepsilon << 1$, we can expect that the model leads to a decelerating expansion
in matter-dominated regime whose existence is also fundamental for the structure formation
process to take place.\\
It is worthwhile to compare our results with recent observations.  To do this, we consider a parametric
approximation of the deceleration parameter along the
cosmic evolution, given by \cite{clarck} \cite{so} \cite{alc}
\begin{equation}
q(z)=q_{0}+q_{1}\frac{z}{z+1}
\label{a21}\end{equation}
The analysis is
performed using the recent SNe Ia observational
data, the so-called Union2 sample of 557 events \cite{a1}.  These observations constrain the parameters $q_{0}$ and $q_{1}$ as follows \cite{clarck} \cite{alc} : for vanishing spatial curvature,
$q_{0}=-0.66 \pm 0.03(1\sigma) \pm 0.07(2\sigma)$ and $q_1=1.54 \pm 0.19(1\sigma) \pm 0.38(2\sigma)$.
In fig.2, we show the evolution
of the deceleration parameter $q(z)$ for the resulting $2\sigma$
intervals of $q_0$ and $q_1$ obtained using the data set considered
above.  As the figure indicates, these observations give a bound on the redshift at which the universe switches from decelerated
to accelerated expansion.  We may use these bounds on the transition redshift to constrain our model parameters.
\begin{figure}[ht]
\begin{center}
\includegraphics[width=0.6\linewidth]{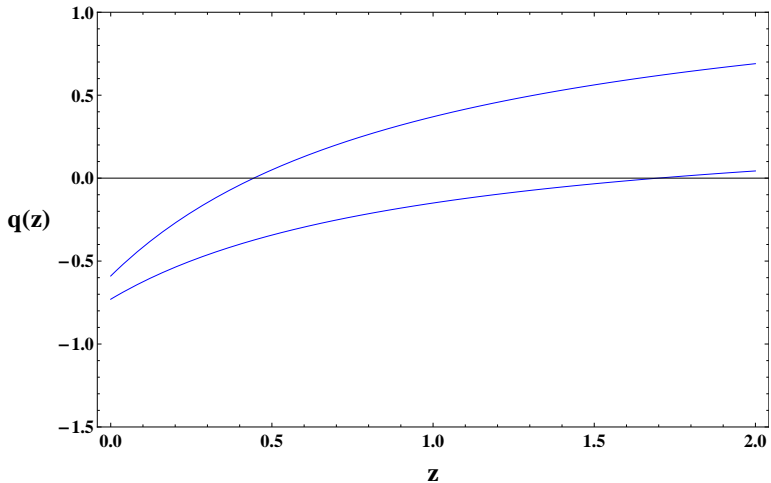}
\caption{Evolution of the deceleration parameter
$q(z)$ for the resulting $2\sigma$ intervals of $q_0$ and $q_1$ obtained using
the data set provided by Union2 sample.}
\end{center}
\end{figure}
However, it should be noted that between the
two parameters $\omega$ and $\beta$ appeared in the expression (\ref{a20}) the former has been already constrained by local gravity experiments.  Thus for
a given $\omega$, (\ref{a20}) gives the deceleration parameter in terms of the redshift and the parameter
$\beta$.  Exploring the equation (\ref{a20}) reveals that, in this case, for a particular value of $\beta$ the transition redshift takes its maximum value.  This behavior is indicated
in fig.3 for $\omega=40000$.  This figure shows that for this value of the BD parameter
the maximum transition redshift lies within the range given by the Union2 sample.  This constrains the parameter $\beta$ to be
up to $0.7$.
\begin{figure}[ht]
\begin{center}
\includegraphics[width=0.6\linewidth]{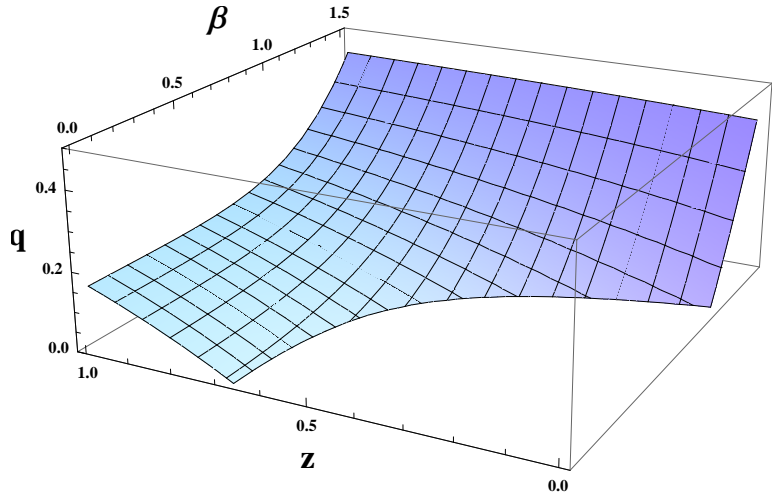}
\caption{The plot of deceleration parameter in terms of $\beta$ and $z$.  }
\end{center}
\end{figure}

In summary, we have considered the possibility that a self-interacting Brans-Dicke field accounts for the
accelerated expansion of the Universe in Einstein conformal frame.  In our analysis, the key point is the ansatz (\ref{a17}) which
has two important features.  First, it provides
a class of solutions of the field equations in terms of a potential of the form (\ref{a18-2}). We have shown that these solutions are consistent
with late-time accelerating expansion of the universe for large values of the BD parameter $\omega$.  Second, it
modifies the evolution of matter density to (\ref{a18}) which is the simplest possible way of stating that
the matter dilution is attenuated due to its interaction with the scalar field $\varphi$.  This evolution law indicates that the deviation from the standard evolution is characterized
by a positive constant parameter $\varepsilon$ which quantifies the decay rate.  Our analysis also indicates that
recent accelerating
expansion is possible for $\varepsilon<<1$ (or $\omega>>1$) in accord with local gravity tests.

\end{document}